\documentclass[epj, nopacs]{svjour}
\usepackage{graphics}
\usepackage{epsfig}
\usepackage{amsmath}
\begin{document}
\title{Generating Functional in String Field Theory}
\author{Am-Gil Ri\inst{1}
\thanks{\emph{Present address:} Daesong District, Pyongyang, DPR Korea}%
 \and Tae-Song Kim\inst{1} \and Song-Jin Im\inst{2}% etc
}                     % Do not remove
\institute{Faculty of Energy Science, Kim Il Sung University \and Faculty of Physics, Kim Il Sung University}
\date{Received: date / Revised version: date}
\abstract{
In our paper, we introduce a path integral of general functional field in order to build the path integral formalism in string field theory from the fact that a string field is a functional field, and describe a method for calculating it in the case of ``Gauss-type". We also obtain the generating functional of an open bosonic string and the corresponding Feynman diagram.
 % end of PACS codes
} %end of abstract
\maketitle
\section{Path integral of a functional field}
\label{sec:1}
Only the operator approach \cite{Siegel} has been used in quantization of string field, but the path integral has not yet.
Feynman’s path integral \cite{Feynman} can be applied to quantization of only point particle field and the Batalin-Vilkovisky formalism \cite{Thorn,Batalin} and Polyakov’s path integral \cite{Polyakov,Hatfield} is mathematically equivalent to Feynman’s, so it can’t be applied to quantization of string field.
 It is necessary to introduce a path integral of a functional field in order to obtain a generating functional of a string field. A string (or superstring) is a particle with one-dimensional structure; therefore, its field is a functional field \cite{Kaku}. Here we’ll deal with a general functional field.

\begin{equation}
\label{eq:1}
\begin{split}
\Phi[x]\equiv&~\Phi(\lbrace x^{\mu}(\sigma ^{m})\mid \sigma ^{m}\in B^{(M)}), \\
	&\begin{pmatrix}
		\mu=&0,1,...,D-1;& \\
                  m=&0,1,...,M-1;&M \le D
	\end{pmatrix}
\end{split}
\end{equation}
, where $D$ is the dimension of the time-space, $M$ is the dimension of the surface, $\sigma^{m}$ are the intrinsic coordinates, and $B^{(M)}$ is a region of the M-dimensional subspace. For example, if $M=1$, it means the field of a point-particle; if $M=2$, of a string; if $M=3$, of a membrane; if $M=4$, of a mass. Then, the action of the field becomes the functional of the functional field.
\begin{equation}
\label{eq:2}
S[\Phi[x]]=\int Dx(\sigma)L(\Phi[x])
\end{equation}
, where $Dx(\sigma)$ is the measure of the path integral and $L(\Phi[x])$ is the Lagrangian density of the functional field.

Now we define the functional derivative of the functional $F[\Phi[x]]$ of the functional field $\Phi[x]$ as follows
\begin{equation}
\label{eq:3}
\frac{\delta F[\Phi[x']]}{\delta\Phi[x]}=\lim_{\varepsilon\rightarrow 0}\frac{1}{\varepsilon}\lbrace  F[\Phi[x']+\varepsilon\delta[x'-x]]-F[\Phi[x]] \rbrace
\end{equation}
, where $\delta[x'-x]$ is called the $\delta$-functional.
Then we'll introduce the path integral of the functional field $\Phi[x]$.
First, we consider the ordinary path integral as follows
\begin{equation}
\label{eq:4}
\begin{split}
I_{N}=&\int D\Phi(x(\sigma_{1}),\cdots,x(\sigma_{N})) \cdot \\
&~~~\cdot F[\Phi(x(\sigma_{1}),\cdots,x(\sigma_{N}))]
\end{split}
\end{equation}
, where $\Phi(x(\sigma_{1}),\cdots,x(\sigma_{N}))$ is the ordinary function with several variables.

Next, if the equation (\ref{eq:4}) has the meaning when $N\rightarrow\infty$ , we call it the path integral of the functional field and define it as
\begin{equation}
\label{eq:5}
\begin{split}
\int{D\Phi[x]F[\Phi[x]]}\equiv \lim_{N\rightarrow\infty}\int D\Phi(x(\sigma_{1}),\cdots,x(\sigma_{N})) \cdot \\ \cdot F[\Phi(x(\sigma_{1}),\cdots,x(\sigma_{N}))]
\end{split}
\end{equation}
, where the measure of the path integral is defined as
\begin{equation}
\label{eq:6}
D\Phi[x]\equiv\lim_{N\rightarrow\infty}\prod_{\lbrace x\rbrace}{d\Phi(x(\sigma_{1}),\cdots,x(\sigma_{N}))}
\end{equation}

If the functional $F[\Phi[x]]$ of the functional field is of ``Gauss-type'', the path integral of the functional field can be calculated explicitly by using the results known in the ordinary path integral and the definition (\ref{eq:5}), i.e.
\begin{equation}
\label{eq:7}
\begin{split}
I[J[x]]&=\int D\Phi[x]\exp\Big\lbrace- \int Dx(\sigma)\cdot \\
& \cdot\Big{(}\frac{1}{2}\Phi[x]A[x]\Phi[x]+
J[x]\Phi[x]\Big{)} \Big\rbrace
\end{split}
\end{equation}
, where $A[x]$ is the operator of the functional. First we calculate $I_{N}$.

\begin{equation*} 
\begin{split}
I_{N}&[J[x]]=\int D\Phi(x(\sigma_{1}),\cdots,x(\sigma_{N})) \exp\Big \lbrace-\int \prod_{i=1}^{N}\\
& dx_{i}\Big\lbrack\frac{1}{2} \Phi(x_{1},\cdots,x_{N})A(x_{1},\cdots,x_{N})\Phi(x_{1},\cdots,x_{N}) \\
&+ J(x_{1},\cdots,x_{N})\Phi(x_{1},\cdots,x_{N}) \Big\rbrack  \Big\rbrace  
\end{split}
\end{equation*}
\begin{equation} % eq.(8)
\label{eq:8}
\begin{split}
~~~=&\left[ \det A(x_{1},\cdots,x_{N})\right]^{-\frac{1}{2}}\exp \Big\lbrace -\frac{1}{2}\prod^{N}_{i=1}dx_{i}dy_{i}\\
&J(x_{1},\cdots,x_{N}) A^{-1}(x_{1},\cdots,x_{N};y_{1},\cdots,y_{N})\\
&\cdot J(y_{1},\cdots,y_{N}) \Big\rbrace
\end{split}
\end{equation}
, where $x_{i}$ means $x(\sigma_{i})$, $\det A(x_{1},\cdots,x_{N})$ is the determinant of the operator $A(x_{1},\cdots,x_{N})$ and $A^{-1}(x_{1},\cdots,x_{N})$ is its inverse operator.
Then if  $N\rightarrow\infty$, $I_{N}\rightarrow I[J[x]]$. Thus we can get the following result.
\begin{equation} %  eq.(9)
\label{eq:9}
\begin{split}
I[J[x]]=&[\det A[x]]^{-1/2}\cdot\\
&~\exp{\Big\lbrace-\int{DxDyJ[x]A^{-1}[x,y]J[y]\Big\rbrace}}
\end{split}
\end{equation}
, where $A^{-1}[x,y]$ is the inverse operator of the functional operator $A[x,y]$.
Also $\det A[x]$, the determinant of $A$, is calculated into
\begin{equation}
\label{eq:10}
\det A[x(\sigma)]=\exp{(tr\ln{A[x(\sigma)]})}=\prod_{j=1}\lambda_{j}(\sigma)
\end{equation}
, where $tr$ means the trace of a operator (i.e. the sum of diagonal elements) and $\lambda_{j}$ are the eigenvalues of $A$ and obtained from the eigenvalue equation
\begin{equation}
\label{eq:eigen}
A[x(\sigma)]\Phi_{j}[x(\sigma)]=\lambda_{j}(\sigma)\Phi_{j}[x(\sigma)].
\end{equation}
If the functional field is the complex scalar field of $\Phi$ and $\Phi^{*}$, the above procedure leads to the following result.
\begin{eqnarray}
\begin{split}
I[J,J^{*}]=& \int D\Phi[x]D\Phi^{*}[x]\exp\Big\lbrace-\int Dx(\sigma) \\ 
&\left(\Phi^{*}[x]A[x]\Phi[x]+J^{*}[x]\Phi[x]+\Phi^{*}[x]J[x]\right)\Big\rbrace
\end{split}\nonumber\\
\label{eq:11}
\begin{split}
&=\left(\det A[x]\right)^{-1}\exp{\Big\lbrace\int{DxDyJ^{*}[x]A^{-1}[x,y]J[y]\Big\rbrace}}
\end{split}
\end{eqnarray}
Then, we introduce a path integral of an anticommutative functional field (Grasman functional field).
The anticommutative functional field is defined as following.
\begin{equation}
\label{eq:12}
\begin{split}
\Psi[x(\sigma)]\equiv&\Psi(\lbrace x^{\mu}(\sigma^{m})\mid \sigma^{m}\in B^{(M)}\rbrace),\\
	&\begin{pmatrix}
		\mu=&0,1,...,D-1;& \\
                  m=&0,1,...,M-1;&M \le D
	\end{pmatrix}
\end{split}
\end{equation}
, where the symbols mean the same as in (\ref{eq:1}).
The functional of an anticommutative field yields the anticommutative number. So $\Psi$ is the generator of Grasman algebra.
Now we'll define the integral of a functional $F[\Psi]$ over $\Psi$ as in the definition (\ref{eq:5}).
\begin{equation}
\label{eq:13}
\begin{split}
\int &D\Psi[x]D\Psi^{+}[x]F[\Psi[x],\Psi^{+}[x]]\equiv\\
&\lim_{N\rightarrow\infty}\int D\Psi(x_{1},\cdots,x_{N})D\Psi^{+}(x_{1},\cdots,x_{N}) \\
&\cdot F[\Psi(x_{1},\cdots,x_{N}),\Psi^{+}(x_{1},\cdots,x_{N})]
\end{split}
\end{equation}

The integral measures in the definition (\ref{eq:13}) can be written like in (\ref{eq:6}) and the same procedures as we got the expression (\ref{eq:9}) enable us to obtain the following result if the functional $F[\Psi]$ is of ``Gauss-type''.
\begin{equation*}
\begin{split}
I[J_{\Psi}[x],&J^{+}_{\Psi}[x]]=\int D\Psi[x]D\Psi^{+}[x] \exp\Big\lbrace-\int Dx(\sigma)\\
                  &(\Psi^{+}[x]A[x]\Psi[x]+J^{+}[x]\Psi[x]+\Psi^{+}[x]J_{\Psi}[x])\Big\rbrace
\end{split}
\end{equation*}
\begin{equation}   % eq.(14)
\label{eq:14}
\begin{split}
~~=(\det A[x])^{-1} \exp&\Big\lbrace-\int DxDy\\
&J^{+}_{\Psi}[x]A^{-1}[x,y]J_{\Psi}[y])\Big\rbrace
\end{split}
\end{equation}
, where the sources $J_{\Psi}$ and $J^{+}_{\Psi}$ are the anticommutative quantities (Grasman elements), too.

\section{The generating functional of a string field}
Now then from the above preparation, we can write the generating functional of the string field theory with the form of the path integral.
For simplicity, we'll consider the boson open string. Its Lagrangian in the light-cone gauge is as the following\cite{Kaku,Cremmer}.
\begin{equation}
\label{eq:15}
\begin{split}
L=& L_{0}+L_{I}  \\
L_{0}=& \int^{\infty}_{0}dp^{+}\int D\vec{x}_{\bot}(\sigma)\Phi^{*}_{p^{+}}[x](i\frac{\partial}{\partial\tau}-H)\Phi_{p^{+}}[x] \\
L_{I}=& \frac{g}{2}\int\prod^{3}_{i=1}\frac{dp^{+}_{i}}{\sqrt{2p^{+}_{i}}}D\vec{x}_{i\bot}(\sigma_{i})\delta(p^{+}_{3}-p^{+}_{1}-p^{+}_{2}) \\
&\cdot\Phi^{*}_{p^{+}_{1}}[x_{1}]\Phi^{*}_{p^{+}_{2}}[x_{2}]\Phi_{p^{+}_{3}}[x_{3}]\delta[x_{1},x_{2},x_{3}]+h.c.
\end{split}
\end{equation}
, where $\vec{x}_{\bot}$ is the transverse component of the vector $\vec{x}$ and note $\delta[x_{1},x_{2},x_{3}]$ is 
\begin{equation} % eq. (16)
\label{eq:16}
\begin{split}
\delta&[x_{1},x_{2},x_{3}]\equiv\prod_{\lbrace\sigma_{3}\rbrace}\delta[\vec{x}_{3\bot}(\sigma_{3})-\vec{x}_{1\bot}(\sigma_{1})\theta_{1}-\vec{x}_{2\bot}(\sigma_{2})\theta_{2}]  \\
&\theta_{1}=\theta(\pi\vert\alpha_{1}\vert-\sigma_{3}),\theta_{2}=\theta(\sigma_{3}-\pi\vert\alpha_{1}\vert) \\
&\alpha=2p^{+}, \alpha_{i}=2p^{+}_{i}, 0<\sigma_{i}<\pi\vert\alpha_{i}\vert 
\end{split}
\end{equation}

In (\ref{eq:15}), Hamiltonian $H$ is as following\cite{Kaku}.
\begin{equation}
\label{eq:17}
H=\frac{\alpha}{2}\pi^{2}\int^{\pi\alpha}_{0}{d\sigma\Big(-\frac{\delta^{2}}{\delta\vec{x}^{2}_{\bot}(\sigma)}+(\frac{\vec{x'}_{\bot}}{2\pi})^{2}\Big)}
\end{equation}

Then the generating functional of the boson string field becomes
\begin{equation}
\label{eq:18}
\begin{split}
Z[J[x],J^{*}[x]]=\int &D\Phi[x]D\Phi^{*}[x]\exp\Big\lbrace  i S[\Phi,\Phi^{*}]\\
&+i\int Dx(\sigma)(J^{*}\Phi+\Phi^{*}J) \Big\rbrace
\end{split}
\end{equation}

This integral is the path integral of the functional field defined above.
The generating functional of the free field is equal to the following.
\begin{equation*}
\begin{split}
Z_{0}[J[x],J^{*}[x]]=&\int D\Phi[x]D\Phi^{*}[x]\exp\Big\lbrace i S_{0}[\Phi,\Phi^{*}]\\
&+i\int{Dx(\sigma)(J^{*}\Phi+\Phi^{*}J)} \Big\rbrace
\end{split}
\end{equation*}
\begin{equation}  %eq. (19)
\label{eq:19}
\begin{split}
~~~~=&\int D\Phi[x]D\Phi^{*}[x]\exp\Big\lbrace i \int d\tau \int_{0}^{\infty}dp^{+}\\
&\int D\vec{x}_{\bot}\Phi^{*}_{p^{+}}[x](i\frac{\partial}{\partial\tau}-H)\Phi_{p^{+}}[x] \Big\rbrace
\end{split}
\end{equation}

This is the path integral of ``Gauss-type'' and so it can be calculated into the following from the equation (\ref{eq:11}).
\begin{equation}
\label{eq:20}
\begin{split}
Z_{0}[J,J^{*}]=&N_{0}\exp\Big\lbrace i\int d\tau d\tau'\int^{\infty}_{0}dp^{+}dp'^{+}\\
&\int D\vec{x}_{\bot}D\vec{x'}_{\bot}(\sigma) J_{p^{+}}(\tau,[\vec{x}_{\bot}]) \\
&\cdot G_{0}(\tau,p^{+},[\vec{x}_{\bot}]\mid \tau',p'^{+},[\vec{x'}_{\bot}]) \\
&\cdot J_{p^{+}}(\tau',[\vec{x'}_{\bot}])
\end{split}
\end{equation}
, where 
\begin{equation}
\label{eq:21}
N_{0}=\Big[\det(i\frac{\partial}{\partial\tau}-H)\Big]^{-1}
\end{equation}
, whereas $G_{0}$ is Green function of the operator $\Big(i\frac{\partial}{\partial\tau}-H\Big)$, i.e.
\begin{equation}
\label{eq:22}
\begin{split}
G_{0}[x,&x']\equiv G_{0}(\tau,p^{+},[\vec{x}_{\bot}]\mid \tau',p'^{+},[\vec{x'}_{\bot}])\\
&=i\theta(\tau-\tau')\delta(p^{+}-p'^{+})B^{-1} \exp\Big\lbrace i \int^{\pi\alpha}_{0}\frac{d\sigma}{\pi\alpha} \\
& ~~~\int^{\pi\alpha}_{0}\frac{d\sigma'}{\pi\alpha}\vec{x}_{i\bot}(\sigma)A_{ij}(\sigma,\tau;\sigma',\tau')\vec{x}_{j\bot}\Big\rbrace\\
&(i,j=1,2)
\end{split}
\end{equation}
, where
\begin{equation}
\label{eq:23}
\begin{split}
\vec{x}_{1\bot}(\sigma)&=\vec{x}_{\bot}(\sigma)\\
\vec{x}_{2\bot}(\sigma)&=\vec{x'}_{\bot}(\sigma)\\
B&=\prod^{\infty}_{n=0}\Big(\frac{i\sin\frac{n\tau}{\alpha}}{n}\Big)^{\frac{D-2}{2}} \\
A_{ij}(\sigma,\tau;\sigma',\tau')&=\sum^{\infty}_{n=0}n
	\begin{pmatrix}
	\displaystyle{\cot\frac{n\tau}{\alpha}} & \displaystyle{\sin^{-1}\frac{n\tau}{\alpha}} \\
	\displaystyle{\sin^{-1}\frac{n\tau}{\alpha}} & \displaystyle{\cot\frac{n\tau}{\alpha}}
	\end{pmatrix}
	\\
	&\qquad\qquad\cos\frac{n\sigma}{\alpha}\cos\frac{n\sigma'}{\alpha}
\end{split}
\end{equation}

Also, the whole generating functional can be calculated by means of perturbation method. From (\ref{eq:18}),
\begin{equation*}
\begin{split}
Z[J,J^{*}]&=\int D\Phi[x]D\Phi^{*}[x]\exp\Big\lbrace i (S_0[\Phi,\Phi^*]+S_I[\Phi,\Phi^*])\\
&\qquad+i\int Dx(\sigma)(J^{*}[x]\Phi[x]+\Phi^{*}[x]J[x])\Big\rbrace\\
&=\exp\Big\lbrace i S_I[\frac{1}{i}\frac{\delta}{\delta J[x]},\frac{1}{i}\frac{\delta}{\delta J^{*}[x]}]\Big\rbrace Z_0[J,J^{*}]
\end{split}
\end{equation*}

Rewrite this, i.e.
\begin{equation}
\label{eq:24}
\begin{split}
Z[J,J^{*}]&=\exp\Big\lbrace i\int d\tau\frac{g}{2}\int\prod^{3}_{i=1}\frac{dp^+_i}{\sqrt{2p^+_i}}Dx_{i\bot}(\sigma)\Big\rbrace \\
&\cdot\delta(p_3^+-p_1^+-p_2^+)\delta[x_1,x_2,x_3])\Big(\frac{1}{i}\frac{\delta}{\delta J_{p^+_1}[x_1]},\\
&\frac{1}{i}\frac{\delta}{\delta J_{p^+_2}[x_2]}\frac{1}{i}\frac{\delta}{\delta J_{p^+_3}[x_3]}+h.c.\Big) Z_0[J,J^{*}]
\end{split}
\end{equation}

From (\ref{eq:24}), we can get the essential quantities (Green function, S-matrix, and so on) of the string field theory. Then we can obtain the Feynman diagram in the time-space.
For example, the perturbation series of Green function of one string can be represented with Feynman diagram as

\begin{figure}[!ht]
\begin{center}
\includegraphics[clip=true,scale=1]{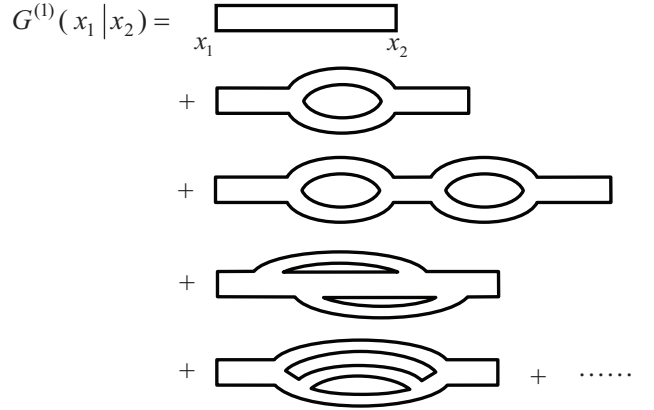}
\caption{\label{fig:ene-vol}Feynman Diagram of perturbation series of Green function of one string}
\end{center}
\end{figure}

\section*{\label{sec:con}Conclusion}
We introduced a functional integral and a functional derivative of a general functional field and described the eigenvalues of a functional operator and the method to calculate the path integral of the general functional field if it is of ``Gauss-type”. We also obtained the generating functional of an open bosonic string field on light cone gauge and calculated the Green function of the free string field, so it is possible to get several necessary quantities of the string field as well as the Green function and the Feynman diagram in space-time.

% BibTeX users please use
% \bibliographystyle{}
% \bibliography{}

\end{document}